\documentclass[12pt,preprint]{aastex}

\newcommand{\etal}{{\it et al.}}

\newcommand{\arcm}{{$^\prime\,$}}
\newcommand{\arcs}{{$^{\prime\prime}\,$}}

\begin{document}

\title{Spurious Shear from the Atmosphere in Ground-Based Weak Lensing
Observations}

\shorttitle{Spurious Shear from the Atmosphere}

\author{D. Wittman\altaffilmark{1}}

\altaffiltext{1}{Physics Department, University of California, Davis,
CA 95616; dwittman@physics.ucdavis.edu}

\begin{abstract}
Weak lensing observations have the potential to be even more powerful
than cosmic microwave background (CMB) observations in constraining
cosmological parameters.  However, the practical limits to weak
lensing observations are not known.  Most theoretical studies of weak
lensing constraints on cosmology assume that the only limits are shot
noise on small scales, and cosmic variance on large scales.  For
future large surveys, shot noise will be so low that other, systematic
errors will likely dominate.  Here we examine a potential source of
additive systematic error for ground-based observations: spurious
power induced by the atmosphere.  We show that this limit will not be
a significant factor even in future massive surveys such as LSST.
\end{abstract}

\section{Introduction}

Weak lensing by large-scale structure imprints correlated
ellipticities onto images of distant galaxies.  This cosmic shear
effect was predicted in the 1960's (Kristian \& Sachs 1966), but not
detected until 2000 (Wittman \etal\ 2000; van Waerbeke \etal\ 2000;
Bacon \etal\ 2000; Kaiser, Wilson \& Luppino 2000).  Cosmic shear can
provide a relatively clean probe of cosmological parameters because it
depends only on the mass distribution and not on detailed astrophysics
such as galaxy bias or gas heating and cooling.  The observational
state of the art has rapidly progressed to quantitative constraints on
dark energy ({\it e.g.} Jarvis \etal\ 2005).  Future large surveys
such as LSST (Tyson \etal\ 2003) and SNAP (Refregier \etal\ 2004) are
intended to make these constraints precise.

However, the practical limits to weak lensing accuracy are not well
known.  The limits which are well known are shot noise on small
scales, and cosmic variance on large scales.  However, shot noise is
so low for future large surveys that other, systematic errors will
likely dominate.  Likely sources of systematics include shear
calibration, photometric redshift errors, spurious power from
instrumental and atmospheric effects, and perhaps intrinsic
alignments.  Some of these have already been addressed in the
literature.  Shear calibration, probably the most important
multiplicative effect, includes a correction for the dilution of shear
by the isotropic smearing of the point-spread function (PSF) by the
atmosphere.  Heymans \etal\ (2005) examined these errors empirically,
by conducting blind analyses of a synthetic dataset, using various
shear calibration methods.  The accuracy of the best current methods
was good to $\sim$1\%.  Guzik \& Bernstein (2005) examined the effect
of spatially varying calibration errors and concluded that limiting
spatial variations in calibration to 3\% rms would be sufficient to
keep systematic errors below statistical errors in future surveys like
LSST.  Ma \etal\ (2005) explored the mitigation of photometric
redshift errors in tomography, and King (2005) did the same for
intrinsic alignments, but neither tried to predict the actual level of
error.

Here we quantify the likely level of an additive systematic in
ground-based observations: spurious power from the atmosphere.  A
fixed realization of atmospheric turbulence imparts position-dependent
ellipticity variations onto the PSF.  If not
removed, this will result in additive spurious power.  With
$\sim$1 PSF star arcmin$^{-2}$ available to diagnose the PSF, and
atmospheric power predicted on smaller scales, this effect surely
cannot be removed completely.  Of course, long exposure times average
over many different realizations of atmospheric turbulence which
should converge to an isotropic PSF (apart from instrumental
aberrations, which can be significant).  One of the goals of this
paper is to measure the potential atmospheric contribution as a
function of exposure time, to aid in the design of future
high-precision surveys.

Every weak lensing analysis has a PSF anisotropy correction.
Therefore the relevant question is not how much spurious ellipticity
is induced by the atmosphere; it is how much remains after the PSF
anisotropy correction.  For that reason, we conduct a mock lensing
analysis of a dense star field rather than rely on atmospheric
modeling.

\section{Dataset}

We take the Large Synoptic Survey Telescope (LSST, Tyson \etal\ 2003)
as a fiducial survey.  The LSST calls for an 8.4 m telescope with a 10
deg$^2$ field of view, repeatedly surveying $\sim$20,000 deg$^2$ of
sky in multiple optical bandpasses ($grizy$).  Each field will be
imaged hundreds of times in some filters, with rather short exposure
times ($\sim$15 seconds).  The motivation for this is two-fold: to
provide time sampling for scientific goals other than weak lensing,
and to provide the ability to ``chop'' the dataset against any desired
variable to identify and remove weak lensing systematics.  

We have identified a real dataset which has many of the parameters
required to explore spurious shear in such a survey: A set of 10- and
30-second exposures of a dense star field taken by the Subaru 8-m
telescope with its prime-focus camera SuprimeCam.  A dense star field
is required to assess the variation of the PSF on small angular
scales; the short exposures and $\sim$0.7\arcs\ FWHM seeing are
typical of LSST; SuprimeCam has the widest field of any 8 m class
telescope/imager; and the 8 m aperture is a good match.  Technically,
this last match may or may not be important.  The expected scaling of
atmosphere-induced ellipticity with aperture size cancels the expected
scaling with exposure time, so that all observations of a given depth
should have a similar level.  However, in practice other factors may
come into play. For example, if the outer scale of atmospheric
turbulence is not much larger than the telescope aperture, the scaling
arguments are invalid.  For this dataset we do not have independent
measurements of the outer scale or any other atmospheric parameters,
but the 8 m aperture will minimize the risk of mismatches with the
LSST dataset.

The exposures were taken on 7 May 2002 through the $i^\prime$ filter,
at airmasses very near unity (1.003---1.012), with seeing slightly
better than 0.7\arcs.  The field was centered at
18:25:59.955 +21:42:19.06, with no dithering.

\section{Analysis}

An anisotropic PSF leads directly to an additive systematic in the
galaxy shape measurements, and thus to spurious shear.  Initial PSF
anisotropy in ground-based images is typically $\sim$5\%.  This
represents a huge ``foreground'' which must be removed to reveal the
$\sim$1\% or smaller cosmic shear signal.  The steps in the PSF
anisotropy correction are: identification of PSF stars; interpolation
of the spatially varying PSF to the position of the galaxy in
question; and correction of the galaxy shape.  

The initial density of stars in these observations is $\sim$8
arcmin$^{-2}$.  We choose a random subset of density 0.9 arcmin$^{-2}$,
typical of current high-latitude surveys, to act as the stars
available for the PSF correction.  The remainder are designated as
test particles to measure the residuals.  These subsets are hereafter
referred to as PSF stars and test stars respectively.  Future surveys
such as LSST may have somewhat more PSF stars available for several
reasons.  First, the angular resolution of ground-based observations
has improved with time, as dome seeing has been minimized and better
sites have been identified at great expense. 
%Some imagers now
%routinely deliver $\sim$0.5\arcs\ FWHM image quality, but these have
%not been used for current weak lensing surveys due to small field of
%view.  
It is a near certainty that LSST seeing will be better than in current
wide-field surveys on older facilities; the LSST plan calls for a
threshold of 0.7\arcs\ for data to be used in the weak lensing
analysis.  This better angular resolution will allow better
star/galaxy separation and a higher density of usable PSF stars.
Second, a large survey can work very hard to identify more stars.  The
survey can perform star-galaxy separation on the best-seeing images
and compile a database of PSF star positions usable in all seeing
conditions, and it can use color information, not just size, to
identify stars.  Neither of these techniques is currently employed.
The following results improve by 20---25\% if the PSF star density is
doubled to 1.8 arcmin$^{-2}$.

The next step is interpolating the PSF to the position of each galaxy
or test star.  Clearly the final results will depend on the
effectiveness of the interpolation scheme.  Typically a polynomial of
order $\sim$3 is fit to each CCD in each exposure.  More sophisticated
schemes have been used to reach smaller scales for fixed patterns
persisting throughout multiple exposures (Jarvis \& Jain 2004), but
the atmosphere is stochastic and is unlikely to be more accurately
diagnosed by such a scheme.  Therefore we use a simple third-order
polynomial.  We also tried nearest neighbor and bicubic spline
interpolation, which fared slightly worse and are not presented here.
The resulting limit on spurious shear is conservative because a better
interpolation scheme may be found.

\begin{figure}
\centerline{\resizebox{3in}{!}{\includegraphics{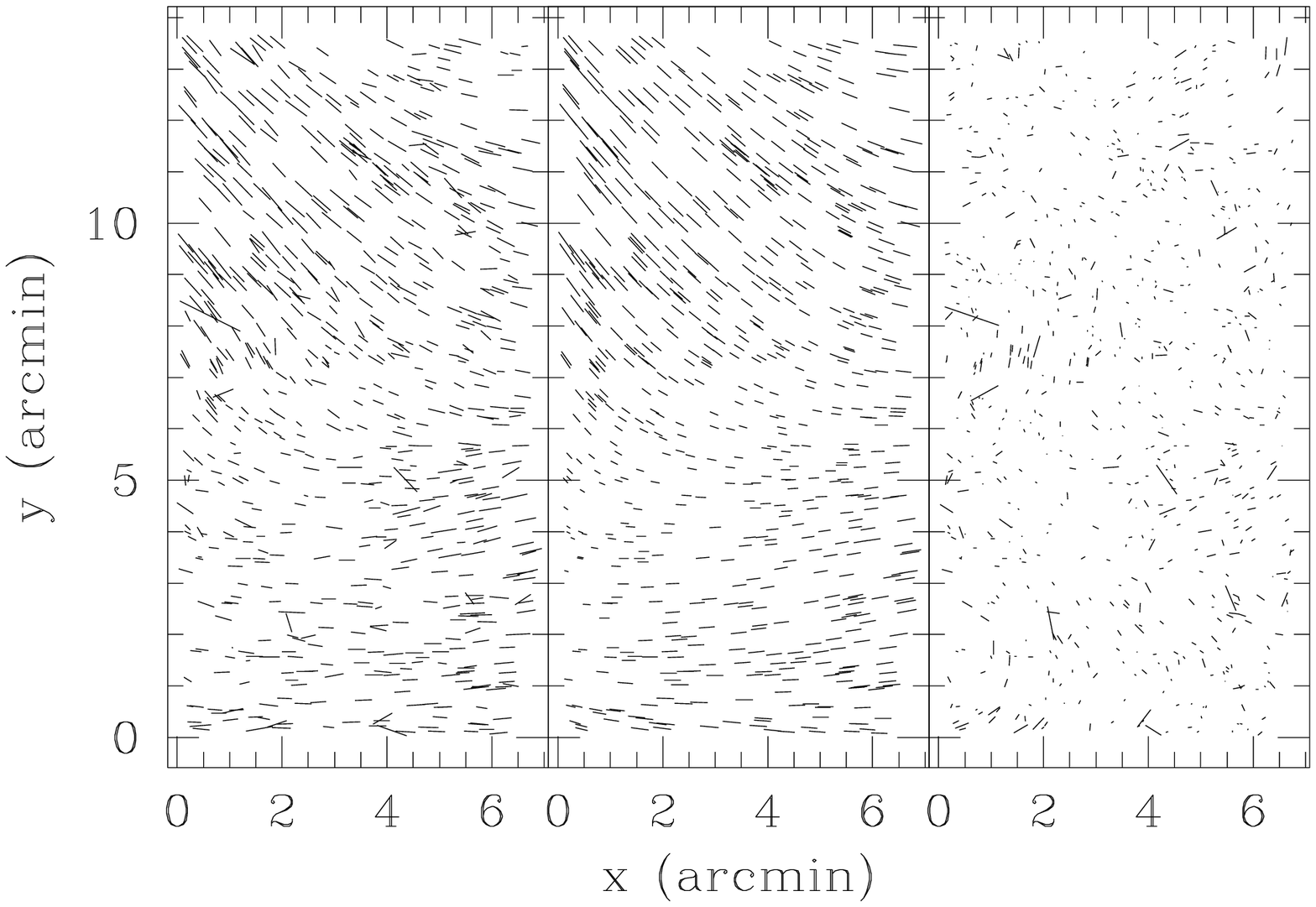}}}
\caption{Left: PSF ellipticity map of one of the ten SuprimeCam CCDs
  in one 10 s exposure, using test stars ($\sim$7
  arcmin$^{-2}$). Center: polynomial model based on disjoint sample of
  $\sim$1 PSF star arcmin$^{-2}$. Right: test star residuals after
  model subtraction.  Apart from blends and other corrupted shape
  measurements, the mean scalar ellipticity before subtraction is 0.07
  and the maximum is 0.12, in the upper left corner.  After
  subtraction, these numbers drop to 0.02 and 0.06.
\label{fig-psf}}
\end{figure}

Figure~\ref{fig-psf} illustrates the interpolation input and output.
The left panel shows the spatial variation of all the measured
ellipticities in one of the ten SuprimeCam CCDs, and the center panel
shows the interpolated ellipticities. At each point where the PSF is
measured or interpolated, we plot a line segment at the PSF position
angle, with length proportional to the PSF ellipticity.  The
atmosphere is not necessarily responsible for all the variation shown.
However, contributions from the optics are likely to be on angular
scales easily modeled by this interpolation process.  Therefore we are
most interested in what remains after removing this pattern.

The final step, galaxy shape correction, also has been implemented in
various ways.  The simplest conceptually is to measure the galaxy
shape after convolving the image with a spatially varying kernel such
that the final PSF is isotropic (Fischer \& Tyson 1997; Kaiser 2000;
Bernstein \& Jarvis 2002).  Here we will assume that the result of the
convolution is as good as the PSF interpolation allows.  This is a
good assumption in practice; and in principle, if the convolution is
imperfect it can be iterated until it is limited only by the PSF
interpolation.  Therefore we subtracted the interpolated ellipticities
componentwise from the measured ellipticities of the test stars,
mimicking a perfectly effective convolution step.  This is illustrated
by the right panel of Figure~\ref{fig-psf}.  Blends and other
corrupted shape measurements now stand out quite clearly.  Because
these have no preferred orientation, they add noise but not bias to
the following measurement.

Finally, we convert the residual ellipticities to shears with the
standard factor of two and compute the shear correlation functions of
the corrected test stars.  In reality, the conversion of PSF
ellipticity to inferred spurious shear depends on the size of the
galaxy relative to the PSF.  If an initially circular source is
resolved, its measured ellipticity will always be less than the PSF
ellipticity.  But for barely-resolved galaxies, the spurious
ellipticity will be amplified by the subsequent correction for
dilution due to the isotropic part of the PSF (variously called seeing
correction, dilution correction, or shear calibration).  These effects
combine to make the inferred pre-seeing ellipticity equal to the PSF
ellipticity for galaxies with pre-seeing size equal to the PSF size;
less for larger galaxies, and more for smaller galaxies.  This
transition size is not atypical for galaxies used in a ground-based
lensing analysis, so we apply no correction factor here, with the
caveat that the impact on a real lensing survey could vary by up to a
factor of $\sim$2, depending on how aggressively the analysis uses
barely-resolved galaxies.

The final result may be an overestimate of the atmospheric
contribution, as there may be remaining unmodelled instrumental
effects.  In a real survey, such effects could be identified in any of
several ways (principal component analysis as in Jarvis \& Jain 2004,
detailed optomechanical modeling of the camera, or measured on the fly
with wavefront sensors) and then removed.

\section{Results}

\begin{figure}
\centerline{\resizebox{3in}{!}{\includegraphics{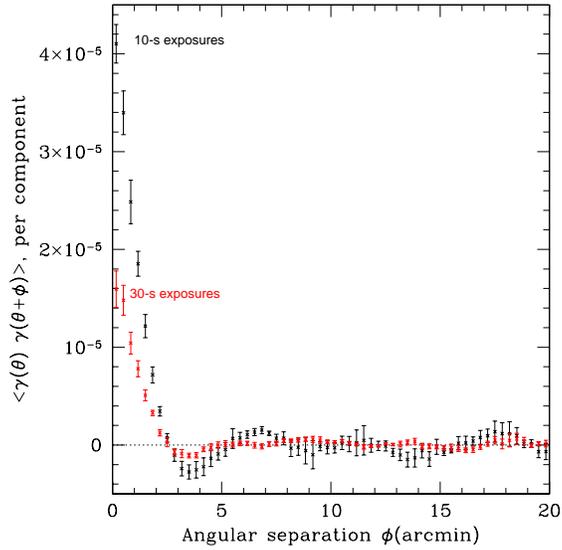}}}
\caption{Residual shear correlations after PSF correction for
10-second (red) and 30-second (black) exposures on the Subaru
telescope.  In each case, only one shear component is plotted for
clarity; the two components are nearly indistinguishable in their
behavior. The ringing at large angular separation is an artifact of
the interpolation scheme.
\label{fig-ocfsingle}}
\end{figure}

As an estimate of the spurious power induced by the atmosphere, we
plot the shear correlations of the corrected test stars in
Figure~\ref{fig-ocfsingle}, for both 10-second (black) and 30-second
(red) exposures.  In each case, only one shear component is plotted
because the two components were indistinguishable.  At each angular
separation, the points and error bars plotted reflect the mean and rms
variation across the five exposures of each duration.  Note that
neighboring points are highly correlated.

\begin{figure}
\centerline{\resizebox{3in}{!}{\includegraphics{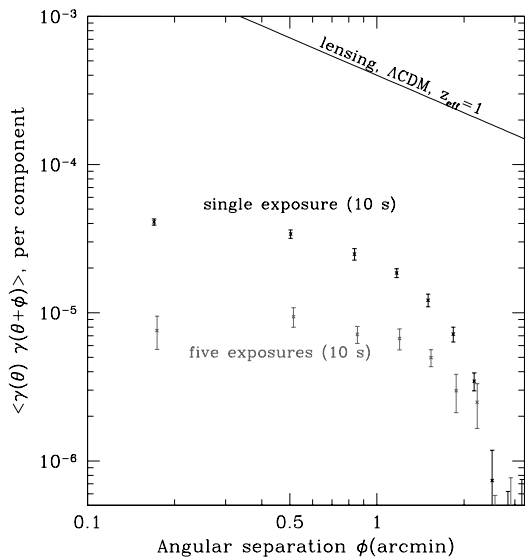}}}
\caption{Decrease of residual shear correlations from a single 10 s
exposure (black) to a coadd of five 10 s exposures (gray, and slightly
offset horizontally for clarity). The expected fivefold decrease is
realized at some, but not all, angular scales, possibly due to
unmodelled instrumental effects.  LSST plans to take 200 exposures in
each filter for each field.  The expected lensing signal is shown for
an effective source redshift of unity and a $\Lambda$CDM universe
(Jain \& Seljak 1997).\label{fig-ocfmulti}}
\end{figure}

For vanishing angular separation, the quantity plotted in
Figure~\ref{fig-ocfsingle} is equivalent to the mean-square value of
the residual atmospheric shear.  For $n$ independent realizations of
atmospheric turbulence, we expect this to go as $n^{-1}$, {\it i.e.},
the rms goes as $n^{-1/2}$.  The improvement from 10-s to 30-s
exposure time is more modest (a factor of 2.4 rather than 3), probably
because the atmosphere has not completely decorrelated in 30 seconds.
Perhaps it would be better to accumulate a longer exposure time by
taking multiple short exposures, alternating fields so that the
atmosphere is completely decorrelated by the time a field is
revisited.  To investigate this possibility, we examined a set of five
consecutive 10-second exposures.  The SuprimeCam read time is long
enough ($\sim$120 seconds) that it is a fair comparison to LSST, with
its fast read time and point/settle time, doing several fields and
coming back for a revisit.  For any reasonable atmosphere, it should
provide complete decorrelation.  For each test star, we took the mean
of the five corrected shapes as its final shape estimate.  The result
is shown in Figure~\ref{fig-ocfmulti}, now zoomed in to small
separations where the correlations are detectable.  The improvement is
indeed a factor of five at the smallest angular scales, but less at
2\arcmin\ scales.  One possible explanation is unmodelled instrumental
effects.  CCD height variations, for example, are expected at scales
some fraction of a CCD size (7\arcm\ in this case), but not at very
small scales, where the atmosphere should dominate.  The effects of
CCD height variations would not average down at all with multiple
undithered exposures, but with sufficient effort they could be calibrated
and removed from a large survey.

The LSST dataset will contain hundreds of visits to each field in each
of two bandpasses which will be observed only in good seeing ($r$ and
$i$).  Therefore the spurious shear correlation from the atmosphere
will be on the order of $10^{-7}$ at 1\arcm\ scales.  This is 3---4
orders of magnitude below the expected lensing shear correlation per
component, which is 4$\times 10^{-4}$ for an effective source redshift of 1.0
in a $\Lambda$CDM universe (Jain \& Seljak 1997).  This is the angular
scale at which the spurious shear is at a maximum relative to the
expected lensing signal, at least in the range of scales measured
here.  The expected shear correlation from lensing does not decline
any more rapidly with angular scale than do these measurements, so
there is little reason to think that spurious shear from the atmosphere
will become a significant factor at any angular scale.  The projected
spurious shear from the atmosphere is also much smaller than the
expected LSST statistical errors, which are as good as $\sim$1\% in
any one redshift bin and angular scale.

\section{Summary and Discussion}

We have measured residual shear correlations in LSST-like short
exposures, after PSF anisotropy correction using current algorithms
and a conservative density of PSF stars.  To the extent that the
standard PSF anisotropy correction models away instrumental effects,
the residual correlations can be interpreted as coming from
atmospheric turbulence.  Because it is unlikely that all instrumental
effects have been modeled perfectly, this can be taken as a
conservative upper estimate of spurious shear from the atmosphere at
unity airmass.  For a real LSST-like survey, airmass effects would
increase the spurious shear by less than a factor of two.  Also,
depending on how aggressively future surveys attempt to use barely
resolved galaxies, the following results could go up or down by a
factor of two.

At 1\arcm\ scales, where the spurious shear is at its maximum relative
to the expected lensing signal, a single 10 s exposure exhibits
residual correlations a factor of $\sim$20 smaller than the lensing
signal.  For longer single exposures up to 30 s, the spurious shear
correlations decrease somewhat more slowly than the inverse of the
exposure time.  For coadds of multiple independent exposures separated
by $\sim$120 s, the correlations at scales $<$1\arcm\ decrease
linearly with the number of exposures, {\it i.e.} the rms spurious
shear goes as $n^{-1/2}$.  At larger scales, the observed decrease is
smaller.  However, with no reason to believe that the temporal nature
of the atmosphere is a function of angular scale, unmodelled
instrumental artifacts at a fixed angular scale must be responsible
for this behavior.  In a real survey, such artifacts could be
diagnosed and modeled out.  In fact, a strategy as simple as dithering
could change this systematic into a random error which would average
down.  With hundreds of independent exposures, the residual
correlations in the coadded LSST dataset will be 3---4 orders of
magnitude less than the signal and comfortably less than the shot
noise.

Spurious shear from the atmosphere will not be a major systematic in
ground-based lensing surveys.  Shear calibration, photometric redshift
errors, and their spatial variations are likely to be more important
sources of systematics.  Heymans \etal\ (2005) showed that the best
current shear calibrations are good to $\sim$1\%.  Overall shear
calibration could be treated as a nuisance parameter in the analysis
with a significant penalty, about a factor of two degradation in the
resulting cosmological parameter errors (Huterer \etal\ 2005).
Spatial variations in the calibration cannot be treated this way even
in principle, and will have to be controlled to $\sim$3\% (Guzik \&
Bernstein 2005), which is probably achievable in future ground-based
surveys.  For photometric redshifts, Ma \etal\ (2005) found that the
bias and scatter in each redshift bin of width 0.1 must be known to
better than about 0.003-0.01 to avoid more than a 50\% increase in
dark energy parameter errors.  This will be a challenge for deep
surveys whose imaging goes beyond the capabilities of other facilities
to provide supporting spectroscopy.  Finally, at small scales
($l>1000$), theoretical uncertainty in predicting the shear power
spectrum due to baryonic effects may be a source of uncertainty at the
$\sim$1\% level (Zhan \& Knox 2004).

Another potentially important systematic is spurious shear due to the
telescope and camera.  Comparisons to current instruments are likely
to be misleading because future surveys will be the first ones built
from the ground up to minimize lensing systematics.  The LSST camera,
for example, will have wavefront sensors throughout the focal plane so
that the exact state of the optics will be known as a function of
time.  Both the pupil and the camera will rotate with respect to the
sky, providing another important way to diagnose and reduce
systematics.  We cannot yet estimate the level of spurious shear due
to the telescope and camera, but because these items are testable in
situ, it seems likely that they will be controlled at least as well as
shear and photometric calibration, which depend on observing
conditions which are not under direct control.  The dataset of
hundreds of exposures of each field will be critical for analyzing the
effects of observing conditions and thus improving the limits on
spatial variations of shear calibration and photometric calibration.

\acknowledgments We thank Vera Margoniner, Garrett Jernigan, Tony
Tyson, Steve Kahn, and John Peterson for valuable discussions.  Based
on data collected at Subaru Telescope and obtained from the SMOKA
science archive at Astronomical Data Analysis Center, which is
operated by the National Astronomical Observatory of Japan.


\begin{thebibliography}{}

\bibitem[Bacon et al.(2000)]{2000MNRAS.318..625B} Bacon, D.~J., Refregier, 
A.~R., \& Ellis, R.~S.\ 2000, \mnras, 318, 625 

\bibitem[Bernstein \& Jarvis(2002)]{2002AJ....123..583B} Bernstein, G.~M., 
\& Jarvis, M.\ 2002, \aj, 123, 583 

\bibitem[Fischer \& Tyson(1997)]{1997AJ....114...14F} Fischer, P., \& 
Tyson, J.~A.\ 1997, \aj, 114, 14 

\bibitem[Guzik \& Bernstein(2005)]{2005astro.ph..7546G} Guzik, J., \& 
Bernstein, G.\ 2005, ArXiv Astrophysics e-prints, arXiv:astro-ph/0507546 

\bibitem[Huterer et al.(2005)]{2005astro.ph..6030H} Huterer, D., Takada, 
M., Bernstein, G., \& Jain, B.\ 2005, ArXiv Astrophysics e-prints, 
arXiv:astro-ph/0506030 
 
\bibitem[Kristian \& Sachs(1966)]{1966ApJ...143..379K} Kristian, J., \& 
Sachs, R.~K.\ 1966, \apj, 143, 379 

\bibitem[Kristian(1967)]{1967ApJ...147..864K} Kristian, J.\ 1967, \apj, 
147, 864 

\bibitem[Heymans et al.(2005)]{2005astro.ph..6112H} Heymans, C., et al.\ 
2005, ArXiv Astrophysics e-prints, arXiv:astro-ph/0506112 

\bibitem[Jain \& Seljak(1997)]{1997ApJ...484..560J} Jain, B., \& Seljak, 
U.\ 1997, \apj, 484, 560 

\bibitem[Jarvis \& Jain(2004)]{2004astro.ph.12234J} Jarvis, M., \& Jain, 
B.\ 2004, ArXiv Astrophysics e-prints, arXiv:astro-ph/0412234 

\bibitem[Jarvis et al.(2005)]{2005astro.ph..2243J} Jarvis, M., Jain, B., 
Bernstein, G., \& Dolney, D.\ 2005, ArXiv Astrophysics e-prints, 
arXiv:astro-ph/0502243 
 
\bibitem[KWL]{KWL} Kaiser, N., Wilson, G. \& Luppino, G. 2000, ArXiv
Astrophysics e-prints, arXiv:astro-ph/0003338

\bibitem[Kaiser(2000)]{2000ApJ...537..555K} Kaiser, N.\ 2000, \apj, 537, 
555

\bibitem[King(2005)]{2005astro.ph..6441K} King, L.\ 2005, ArXiv 
Astrophysics e-prints, arXiv:astro-ph/0506441 
 
\bibitem[Ma et al.(2005)]{2005astro.ph..6614M} Ma, Z., Hu, W., \& Huterer, 
D.\ 2005, ArXiv Astrophysics e-prints, arXiv:astro-ph/0506614 

%\bibitem[Margoniner et al.(2004)]{2004AAS...20514803M} Margoniner, V., 
%Wittman, D., Tyson, T., \& Deep Lens Survey 2004, American Astronomical 
%Society Meeting Abstracts, 205, 14803

\bibitem[Refregier et al.(2004)]{2004AJ....127.3102R} Refregier, A., et 
al.\ 2004, \aj, 127, 3102 
 
\bibitem[Tyson et al.(2003)]{2003NuPhS.124...21T} Tyson, J.~A., Wittman, 
D.~M., Hennawi, J.~F., \& Spergel, D.~N.\ 2003, Nuclear Physics B 
Proceedings Supplements, 124, 21 
 
\bibitem[Van Waerbeke et al.(2000)]{2000A&A...358...30V} Van Waerbeke, L., 
et al.\ 2000, \aap, 358, 30 

\bibitem[Wittman et al.(2000)]{2000Natur.405..143W} Wittman, D.~M., Tyson, 
J.~A., Kirkman, D., Dell'Antonio, I., \& Bernstein, G.\ 2000, \nat, 405, 
143 

\bibitem[Zhan \& Knox(2004)]{2004ApJ...616L..75Z} Zhan, H., \& Knox, L.\ 
2004, \apjl, 616, L75 
 

\end{thebibliography}
\end{document}